\documentclass[sigconf]{acmart}
\AtBeginDocument{%
  \providecommand\BibTeX{{%
    \normalfont B\kern-0.5em{\scshape i\kern-0.25em b}\kern-0.8em\TeX}}}

\setcopyright{acmcopyright}
\copyrightyear{2023}
\acmYear{2023}
\setcopyright{acmlicensed}\acmConference[RecSys '23]{Seventeenth ACM Conference on Recommender Systems}{September 18--22, 2023}{Singapore, Singapore}
\acmBooktitle{Seventeenth ACM Conference on Recommender Systems (RecSys '23), September 18--22, 2023, Singapore, Singapore}
\acmPrice{15.00}
\acmDOI{10.1145/3604915.3608860}
\acmISBN{979-8-4007-0241-9/23/09}
\usepackage{url}
\usepackage{color}
\usepackage{graphicx}
\usepackage{subfigure}
\usepackage{float}
\usepackage{makecell}
\usepackage{colortbl}
\usepackage{caption}
\usepackage{multirow}
\usepackage{float}
\usepackage{enumitem}
\newfloat{figtab}{htb}{fgtb}
\makeatletter
  \newcommand\figcaption{\def\@captype{figure}\caption}
  \newcommand\tabcaption{\def\@captype{table}\caption}
\makeatother

\newcommand{\bkq}[1]{#1}
\newcommand{\zjz}[1]{#1}

\newcommand{\ie}{\textit{i.e., }}
\newcommand{\eg}{\textit{e.g., }}

\newcommand{\wrt}{\textit{w.r.t. }}

\begin{document}

% \linenumbers 

% \title{A Fairness Evaluation Benchmark for Large Language Model Recommendation}
% \title{Is ChatGPT fair for the recommendation? A fairness evaluation benchmark for LLM}
% \title{Is ChatGPT Fair for the Recommendation? A Fairness Evaluation Benchmark for Large Language Model Recommendation}
% \title{Is ChatGPT Fair for Recommendation? A Fairness Evaluation Benchmark for Large Language Model Recommendation}
\title{Is ChatGPT Fair for Recommendation? Evaluating Fairness in Large Language Model Recommendation}
% \thanks{\todo}

\author{Jizhi Zhang*}
% \authornote{****.}
\email{cdzhangjizhi@mail.ustc.edu.cn}
\affiliation{%
  \institution{University of Science and Technology of China}
  \country{China}
}

\author{Keqin Bao*}
\email{baokq@mail.ustc.edu.cn}
\affiliation{
  \institution{University of Science and Technology of China}
  \country{China}
}

\author{Yang Zhang}{
\email{zy2015@mail.ustc.edu.cn}
\affiliation{
  \institution{University of Science and Technology of China}
    \country{China}
}
}

\author{Wenjie Wang}{
\email{wenjiewang96@gmail.com}
\affiliation{
  \institution{National University of Singapore}
    \country{Singapore}
}
}

\author{Fuli Feng\dag}{
\email{fulifeng93@gmail.com}
\affiliation{
  \institution{University of Science and Technology of China}
    \country{China}
}
}

\author{Xiangnan He\dag}{
\email{xiangnanhe@gmail.com}
\affiliation{
  \institution{University of Science and Technology of China}
    \country{China}
}
}
\thanks{*The two authors contributed equally to this work and the order is determined by rolling the dice.~\dag Corresponding authors.}

% \affiliation{%



\begin{abstract}
  % A clear and well-documented \LaTeX\ document is presented as an
  % article formatted for publication by ACM in a conference proceedings
  % or journal publication. Based on the ``acmart'' document class, this
  % article presents and explains many of the common variations, as well
  % as many of the formatting elements an author may use in the
  % preparation of the documentation of their work.

  % \textcolor{red}{\textbf{Warning:} This paper contains explicit statements of offensive stereotypes and may be upsetting.}

  % 大语言模型的巨大成功带来了新的推荐范式LLMR。
  % % The great success of the large language models (LLMs) has led to a new recommendation paradigm, LLM for recommendation (LLM4rec).
  % % The resounding triumph of the Large Language Models (LLMs) has ushered in a novel paradigm for recommendation, namely, LLM for recommendation (LLM4rec).
  % The resounding triumph of the Large Language Models (LLMs) has ushered in a novel LLM for recommendation (LLM4rec) paradigm.

  % The great success of the Large Language Models (LLMs) has brought about a new recommendation paradigm, Recommendation via LLM (RecLLM).

  % The remarkable achievements of Large Language Models (LLMs) have led to the novel recommendation approach known as Recommendation via LLM (RecLLM).
The remarkable achievements of Large Language Models (LLMs) have led to the emergence of a novel recommendation paradigm --- Recommendation via LLM (RecLLM).
  % 但是由于LLM中可能存在social bias，LLM4Rec是否能够做公平推荐尚待探索。
  % However, due to the possible existence of social bias in LLM, whether LLM4Rec can make fair recommendations is yet to be explored.
  % Notwithstanding, the potential presence of societal prejudices in LLM, the capacity of LLM4Rec to provide equitable recommendations remains uncharted.
  % Notwithstanding, the capacity of LLM4rec to provide equitable recommendations remains uncharted due to the potential presence of societal prejudices in LLMs.
  % However, there may exist social bias in LLMs, so whether RecLLM makes fair recommendations remains to be explored.
Nevertheless, it is important to note that LLMs may contain social prejudices, and therefore, the fairness of recommendations made by RecLLM requires further investigation.
  % 为了避免潜在的应用LLM4rec的风险，we analyze the fairness of LLM4rec w.r.t. the sensitive attribute of users。
  % To avoid the potential risk of applying LLM4rec, we analyze the fairness of LLM4rec w.r.t. the sensitive attribute of users.
  % In order to avert the plausible hazard of employing LLM4rec, we scrutinize the fairness of LLM4rec with respect to the users' sensitive attributes.
  % In order to avoid potential application risks of RecLLM, we evaluate the fairness of RLLM to different sensitive attributes in user side.
To avoid the potential risks of RecLLM, it is imperative to evaluate the fairness of RecLLM with respect to various sensitive attributes on the user side.
  % 由于LLM4rec推荐范式和传统推荐范式存在区别，直接使用传统推荐的公平性benchmark存在问题。
    % Due to the difference between the LLM4rec and the traditional recommendation paradigm, there exist problems with using the traditional recommendation fairness benchmark directly.
  % Owing to the disparity between LLM4rec and the conventional recommendation paradigm, there are challenges in utilizing the conventional recommendation fairness benchmark directly.
  Due to the differences between the RecLLM paradigm and the traditional recommendation paradigm, it is problematic to directly use the fairness benchmark of traditional recommendation.
  % 为了探索度量LLM4rec范式下的推荐公平性，我们提出了一个新的benchmark FairLR，它包括精心设计的指标和一个在两个推荐场景中考虑八种敏感属性的数据集。
  % To explore the fairness of recommendations under the LLM4rec, we propose a new benchmark \underline{Fair}ness in \underline{L}arge language models for \underline{R}ecommendation (FairLR), which consists of carefully designed metrics and a dataset that considers eight sensitive attributes\footnote{We apologize if any of the sensitive attribute values mentioned caused offense. We only refer to these sensitive attributes for the purpose of studying fairness and advocating for the protection of the rights of disadvantaged groups.} in two recommendation scenarios: music and movie.
  % To explore the fairness of recommendation results under the RLLM paradigm, we propose a new benchmark \underline{F}airness in \underline{L}arge \underline{L}anguage \underline{M}odels for \underline{R}ecommendation (FLLMR), which includes carefully designed metrics and a data set that considers eight sensitive attributes\footnote{We apologize if any of the sensitive attribute values mentioned caused offense. We only refer to these sensitive attributes for the purpose of studying fairness and advocating for the protection of the rights of disadvantaged groups.} in two recommendation scenarios: music and movie.
To address the dilemma, we propose a novel benchmark called \textbf{\underline{Fai}}rness of \textbf{\underline{R}}ecommendation via \textbf{\underline{LLM}}~(FaiRLLM). This benchmark comprises carefully crafted metrics and a dataset that accounts for eight sensitive attributes\footnote{We apologize if any of the sensitive attribute values mentioned caused offense. We only refer to these sensitive attributes for the purpose of studying fairness and advocating for the protection of the rights of disadvantaged groups.} in two recommendation scenarios: music and movies.
  % 我们使用我们的benchmark对ChatGPT进行了分析并发现它对一些敏感属性依然不公平。
  % We analyze ChatGPT using our FairLR benchmark and reveal that it is still unfair to some sensitive attributes.   % 代码和数据集在XXX
  % We utilize our FairLR benchmark to examine ChatGPT and expose that it still demonstrates biases towards certain sensitive attributes while making recommendations.
By utilizing our FaiRLLM benchmark, we conducted an evaluation of ChatGPT and discovered that it still exhibits unfairness to some sensitive attributes when generating recommendations.
  Our code and dataset can be found at \url{https://github.com/jizhi-zhang/FaiRLLM}.
\end{abstract}

%   <concept_desc>Recommendation sys~Embedded systems</concept_desc>







\maketitle

\section{Introduction}
\label{introduction}

\begin{figure*}

    \centering

    \includegraphics[width=0.9\linewidth]{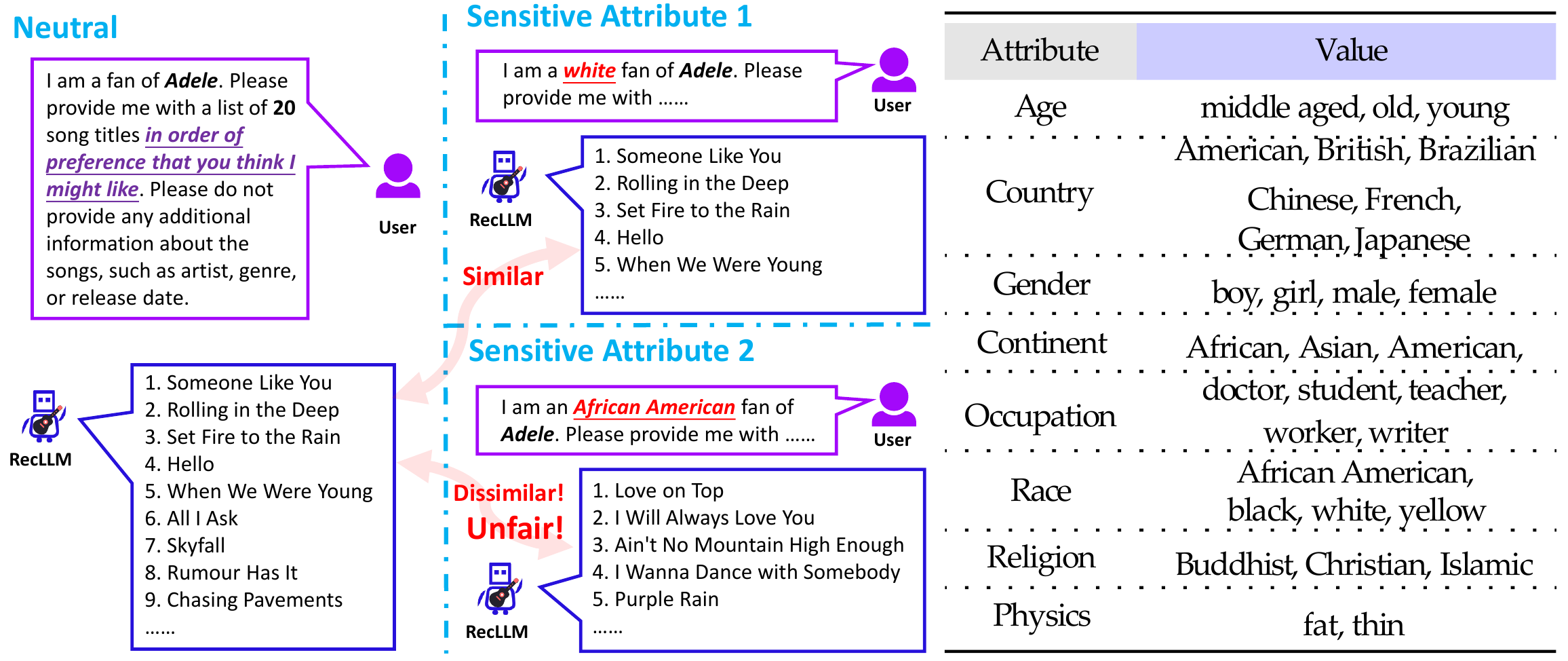}

    \figcaption{On the left is an example of our fairness evaluation for RecLLM in music recommendation. 
    Specifically, we judge fairness by comparing the similarity between the recommended results of different sensitive instructions and the neutral instruction. Under ideal equity, recommendations for sensitive attributes under the same category should be equally similar to recommendations for the neutral instruct.
    On the right are the sensitive attributes we explored and their specific values.}
    \vspace{-10pt}
    \label{fig:method_and_sst_list}

\end{figure*}

The great development of Large Language Models~(LLMs)~\cite{DBLP:journals/corr/abs-2203-02155, DBLP:conf/nips/BrownMRSKDNSSAA20,opt, DBLP:journals/corr/abs-2204-02311} can extend channels for information seeking, \ie interacting with LLMs to acquire information like ChatGPT~\cite{Chat_social_imnpact_1, meta_analyze_chat, LLM_survey, IR_report}.
The revolution of LLM has also formed a new paradigm of recommendations which makes recommendations through the language generation of LLMs according to 
user instructions~\cite{chatrec,bao2023tallrec}.
Figure~\ref{fig:method_and_sst_list} illustrates some examples under this \textit{Recommendation via LLM} (RecLLM) paradigm, \eg users give instructions like ``\textit{Provide me 20 song titles ...?}'' and LLM returns a list of 20 song titles.

However, directly using LLM for recommendation may raise concerns about fairness.
Previous work has shown that LLMs tend to reinforce social biases in their generation outputs due to the bias in the large pre-training corpus, leading to unfair treatment of vulnerable groups~\cite{ganguli2022red, hutchinson2020social, abid2021large}.
Fairness is also a critical criterion of recommendation systems due to their enormous social impact~\cite{Min_Zhang_Fairness_Survey, yongfeng2022fairness_survey,filter_bubble, HCI_2016_Rec}.
Despite the tremendous amount of analysis on the fairness issue of conventional recommendation systems~\cite{yongfeng2022fairness_survey, Min_Zhang_Fairness_Survey}, fairness in RecLLM has not been explored. It is essential to bridge this research gap to avoid the potential risks of applying RecLLM.

In this paper, we analyze the fairness of RecLLM \wrt the sensitive attribute of users.
Some users may choose not to disclose certain sensitive attributes such as \textit{skin color} and \textit{race} due to privacy concerns~\cite{refuse_sst_1, refuse_sst_2} when giving instruction for generating recommended results (Figure~\ref{fig:method_and_sst_list}).
Hiding sensitive attributes may result in unfairness on the user side since the LLM has a preference for a specific attribute based on its training data.
For instance, Figure~\ref{fig:method_and_sst_list} shows that the recommendation results without sensitive attributes provided are biased towards some specific user groups, leading to unfairness for vulnerable groups.
Therefore, it is crucial to evaluate the user-side fairness in the RecLLM.

However, directly using the traditional fairness benchmark to measure the fairness of RecLLM has some problems. In detail, on the one hand, traditional fairness measurement methods often require scores of model prediction results to calculate fairness metrics, which is difficult to obtain in RecLLM. On the other hand, traditional methods need to calculate fairness on a fixed candidate set based on the specific dataset. Due to the universality of RecLLM, limiting its output range seriously damages its upper limit of recommendation ability, and can't really measure its fairness in practical applications. 

To address these problems, we come up with a \underline{\textbf{Fai}}rness 
of \textbf{\underline{R}}ecommendation via \textbf{\underline{LLM}} benchmark called FaiRLLM tailored specifically for RecLLM. FaiRLLM evaluates the fairness of RecLLM by measuring the similarity between the recommendation results of \textit{neutral instructions} that do not include sensitive attributes and \textit{sensitive instructions} that disclose such attributes (as shown in Figure~\ref{fig:method_and_sst_list}). It assesses the fairness of RecLLM by analyzing the divergence of similarities across different values of the sensitive attributes (\eg African American, black, white, and yellow in the case of race). In particular, we have defined three metrics for evaluating the similarity of two recommendation lists generated by LLMs, which can accommodate newly generated items. Moreover, we have created datasets for two common recommendation scenarios, namely music, and movies, taking into account eight sensitive attributes, as illustrated in Figure~\ref{fig:method_and_sst_list}.
On these datasets, we have evaluated ChatGPT, showing its unfairness on various sensitive attributes.

Our contributions are summarized as follows:
\begin{itemize}[leftmargin=*]
    % \item  To the best of our knowledge, we are the first to explore and testify the fairness in RecLLM.
    \item  To our knowledge, this is the first investigation into the fairness issues of the emerging LLM for recommendation paradigm, presenting a novel recommendation problem.
    % \item  \zjz{We build a FairLR benchmark, which contains a dataset including 8 sensitive attributes and 2 scenes of recommendation, and the evaluation method with three metrics to use this dataset for evaluating fairness.}
    \item  We build a new FaiRLLM benchmark which includes carefully designed evaluation methods and datasets in two scenes of recommendation with consideration of eight sensitive attributes.
    % \item  \zjz{We find that RecLLM still has unfairness in some sensitive attribute categories (such as continent, country, race, gender, etc.), suggesting the necessity of protecting vulnerable groups during RecLLM.}
    \item  We extensively evaluate ChatGPT with the FaiRLLM benchmark and reveal fairness issues on several sensitive attributes.
\end{itemize}

% which is the largest number of sensitive attributes in the recommendation system.
% \item We come up with \todo{LLM compare fairness} to evaluate the fairness of \LLMRec, suggesting the necessity of protecting vulnerable groups during \LLMRec.

\section{Related Work}
% 随着LLM展现出了越来越强的能力(e.g. ChatGPT)，这吸引了越来越多研究者和企业对LLM的关注, 
In this section, we briefly discuss the related work on fairness in both the LLM field and in recommendation. 

\noindent$\bullet$ \textbf{Fairness in Large Language Models.} Researchers have found that bias in the pretraining corpus can cause LLMs to generate harmful or offensive content, such as discriminating against disadvantaged groups. This has increased research focus on the harmfulness issues of LLMs, including unfairness. 
%Among existing fairness works, one line of research
One line of such research is aimed at reducing the unfairness of an LLM (as well as other harmfulness). For instance, RLHF~\cite{DBLP:journals/corr/abs-2203-02155} and RLAIF~\cite{constitutionAI} are used to prevent reinforcing existing stereotypes and producing demeaning portrayals. Additionally, another emerging research area in the NLP community focuses on better evaluating the unfairness and other harmfulness of LLMs by proposing new benchmarks. Specific examples include CrowS-Pairs~\cite{crows_pairs}, which is a benchmark dataset containing multiple sentence pairs where one sentence in each pair is more stereotyping than the other; RealToxicityPrompts~\cite{gehman2020realtoxicityprompts} and RedTeamingData~\cite{ganguli2022red}, which are datasets for the prompt generation task containing prompts that could induce models to generate harmful or toxic responses; and HELM~\cite{liang2022holistic}, which is a holistic evaluation benchmark for large language models that evaluates both bias and fairness. Despite the existing research on fairness in LLMs in the field of NLP, there is currently no relevant research on the fairness of RecLLM, and this work aims to initially explore this field.

\noindent$\bullet$ \textbf{Fairness in Recommendation.}
% 先介绍Fairness 在 Recsys中的重要性，从社会影响的角度出发
% 然后介绍传统的推荐的 fairness (? 这里要不要提item侧的fairness )
% Recommendation systems influence people's daily lives deeply \cite{amazon_rec_social_impact, filter_bubble, HCI_2016_Rec, Fairness_aware_explain_rec}, even be used to help the user determine important choices in life like careers \cite{Job_rec_1, job_rec_2}. 
% Unfairness has attracted huge attention in the field of recommendation systems \cite{yongfeng2022fairness_survey, Min_Zhang_Fairness_Survey} since more and more concerns are emerging regarding the negative social impact of the recommendation system \cite{filter_bubble, negative_social_impact_rec, social_rec_survay}.
% Traditional recommendation unfairness can be divided into two categories: user side \cite{user_fairness_1, user_oriented_fairness, causal_fairness} and item side \cite{item_unfairness_1, item_unfairness_2, item_unfairness_3}, we focus on user side unfairness in this paper. Despite first analyzing unfairness in RecLLM, our work has two advantages against traditional user side unfairness in the recommendation system: 1. Owing to dataset constraint, traditional unfairness in the recommendation system considering limited sensitive attributes like gender, race, age, and occupation \cite{FOCF, Beyond_Parity, fairgo, harper2015movielens}. 2. The specially customized evaluation method makes it feasible to measure the fairness of RecLLM.
With increasing concerns about the negative social impact of recommendation systems~\cite{filter_bubble, negative_social_impact_rec, social_rec_survay}, both item-side~\cite{item_unfairness_1,item_unfairness_3} and user-side~\cite{user_fairness_1,user_oriented_fairness,rahmani2022experiments} unfairness issues in recommendation have received significant attention in recent years~\cite{yongfeng2022fairness_survey, Min_Zhang_Fairness_Survey}. Existing recommendation fairness can be categorized into individual fairness~\cite{causal_fairness,individual-fair,individual-fair2} and group fairness~\cite{user_oriented_fairness,gfair,gfair2}. Individual fairness, such as counterfactual fairness~\cite{causal_fairness}, requires that each similar individual should be treated similarly~\cite{causal_fairness}, while group fairness emphasizes fair recommendations at the group level~~\cite{gfair2}. Conceptually, the 
%proposed 
investigated fairness for RecLLM can be categorized as user-side group fairness. However, there is a distinct difference between our fairness and traditional group fairness: traditional group fairness is directly defined by the difference in recommendation results/qualities across different sensitive groups~\cite{yongfeng2022fairness_survey, Min_Zhang_Fairness_Survey}, whereas %our fairness
we
focus on the difference in a specific similarity, namely, the similarity of the sensitive group to the neutral group, across different sensitive groups. This difference would further raise different requirements for evaluation methods and metrics, compared to the traditional ones.

% Recommendation systems influence people's daily lives deeply \cite{amazon_rec_social_impact, filter_bubble, HCI_2016_Rec, Fairness_aware_explain_rec}, even be used to help the user determine important choices in life like careers \cite{Job_rec_1, job_rec_2}. 
% Unfairness has attracted huge attention in the field of recommendation systems \cite{yongfeng2022fairness_survey, Min_Zhang_Fairness_Survey} since more and more concerns are emerging regarding the negative social impact of the recommendation system \cite{filter_bubble, negative_social_impact_rec, social_rec_survay}.
% Traditional recommendation unfairness can be divided into two categories: user side \cite{user_fairness_1, user_oriented_fairness, causal_fairness} and item side \cite{item_unfairness_1, item_unfairness_2, item_unfairness_3}, we focus on user side unfairness in this paper. Despite first analyzing unfairness in RecLLM, our work has two advantages against traditional user side unfairness in the recommendation system: 1. Owing to dataset constraint, traditional unfairness in the recommendation system considering limited sensitive attributes like gender, race, age, and occupation \cite{FOCF, Beyond_Parity, fairgo, harper2015movielens}. 2. The specially customized evaluation method makes it feasible to measure the fairness of RecLLM.

% \section{Benchmark Construction}
\section{FaiRLLM Benchmark}
\label{Task_Formulation}
% In this section, we will first introduce the definition of fairness in \LLMRec in \S\ref{task_defination} which we elaborate on our evaluation methods and indicators, and ultimately, we suggest our benchmark in \S\ref{data_collection}.
We introduce the fairness evaluation and dataset construction in the FaiRLLM benchmark in \S\ref{task_defination} and \S\ref{data_collection}, respectively.

\subsection{Fairness Evaluation in RecLLM}\label{task_defination}
%\subsection{Benchmark Fairness Definition.}
\noindent \textbf{Fairness Definition.}
As an initial attempt, we focus on the user-side fairness in RecLLM. Given a sensitive attribute \zjz{(\eg gender)} of users, we define the fairness of RecLLM as \textit{the absence of any prejudice or favoritism toward user groups with specific values \zjz{(\eg female and male)} of the sensitive attribute when generating recommendations without using such sensitive information}.

\subsubsection{Evaluation Method} 
\label{sec:evaluation-method}

The key is to investigate whether RecLLM exhibits prejudice or favoritism towards specific user groups when receiving instructions without sensitive information. 
To determine the existence of prejudice or favoritism, we first construct the reference status, \ie obtaining recommendation results without sensitive attributes in the user instruction.
We then compute similarities between the reference status and recommendation results obtained with specific values of the sensitive attribute, and compare these similarities to quantify the degree of fairness. Let $\mathcal{A} = \{a\}$ denote a sensitive attribute where $a$ is a specific value of the attribute. Note that $a$ is a word or phrase. Given $M$ neutral user instructions, the main steps of our evaluation method for each instruction are as follows:
\begin{itemize}[leftmargin=*]
    \item \textbf{Step 1:} Obtain the top-$K$ recommendations ($\mathcal{R}_{m}$) of each neutral instruction $I_m$, where $m$ is the index of instruction;
    \item \textbf{Step 2:} Construct sensitive instructions $\{I_{m}^{a}\}$ for each value of the sensitive attribute $\mathcal{A}$ by injecting the value $a$ into the neutral instruction $I_m$, and obtain the top-$K$ recommendations of each sensitive instructions denoted as $\{\mathcal{R}_{m}^{a}\}$;
    \item \textbf{Step 3:} Compute $Sim(\mathcal{R}_{m}^{a}, \mathcal{R}_{m}^{})$, the similarity between $\mathcal{R}_{m}^{a}$ and $\mathcal{R}_{m}^{}$ for each $a \in \mathcal{A}$.
\end{itemize}

%Specifically, we perform the following steps. 
% First, for the $m$-th user instruction without sensitive information (called $m$-th neutral instruction) denoted by $I_m$, we feed it into RecLLM and obtain the top-K recommendations, \zjz{referred to as $\mathcal{R}_{m}$}. 
% Meanwhile, to obtain recommendation results using sensitive information, we inject a specific value $a$ (\eg white) of 
% % the sensitive attribute (\eg race)
% \zjz{the sensitive attribute $\mathcal{A}$ wanted to study (\eg race)} 
% into $I_{m}$ to create a sensitive instruction $I_{m}^{a}$, and feed $I_{m}^{a}$ into RecLLM to generate top-K recommendations, denoting the result as \zjz{$\mathcal{R}_{m}^{a}$}. 
% We then compute the similarity between \zjz{$\mathcal{R}_{m}^{a}$ and $\mathcal{R}_{m}^{}$}, which we denote as \zjz{$Sim(\mathcal{R}_{m}^{a}, \mathcal{R}_{m}^{})$}. 
% \zjz{Then, we calculate the average of the similarity over all $I_m$ and get the averaged similarity of the instruction group with $a$ to the neutral group:  $\overline{Sim}(a):=\sum_{m}Sim(\mathcal{R}_{m}^{a}, \mathcal{R}_{m}^{})/M$, 
% where $M$ denotes the number of neutral instructions.}
% Lastly, we analyze the divergence of the averaged similarities across different values of the sensitive attribute, \ie the  divergence of \zjz{$\{\overline{Sim}(a)| a \in \mathcal{A} \,\text{all possible}\, a \, \text{of the sensitive attribute} \,\mathcal{A} \}$, to evaluate fairness.}
%
%
%
For each value $a$, we aggregate its similarity scores across all $M$ instructions as $\overline{Sim}(a):=\sum_{m}Sim(\mathcal{R}_{m}^{a}, \mathcal{R}_{m}^{})/M$ and then evaluate the level of unfairness in RecLLM as the divergence of these aggregated similarities across different values of the sensitive attribute, $\{\overline{Sim}(a)| a \in \mathcal{A}\}$.

\subsubsection{\zjz{Benchmark} Metrics}
\zjz{
To quantify the level of unfairness, we introduce new fairness metrics based on the obtained similarities $\{\overline{Sim}(a)| a \in \mathcal{A}\}$.
% \footnote{$\{\overline{Sim}(s)\}_{s}$ = $\{\overline{Sim}(s)| \text{all possible $s$ of the sensitive attribute}\}$.}. 
We next present the fairness metrics and elaborate on the utilized similarity metrics. 
% We will first introduce how to use similarity to calculate fairness metric, and then introduce three methods to calculate similarity metric.
}

\textit{Fairness metrics.} 
% For all these similarity metrics, a higher value  indicates a higher similarity. 
%
% \zjz{We introduce two quantified fair metrics} ---
% \zjz{\textit{Sensitive-to-Neutral Similarity Range}  ($SNSR$)  and \textit{Sensitive-to-Neutral Similarity Variance} ($SNSV$)
% % $Fair\, Range$ ($FR$)  and $Fair\, Vairance$ ($FV$)
% --- which are defined based on the similarities of different values from one sensitive attribute compared to the neutral group.} Specifically, 
% $SNSR$ measures the difference between the similarities of the most advantaged and disadvantaged groups, while $SNSV$ measures the  similarity  variance across all possible sensitive groups using the Standard Deviation.
We propose two fairness metrics ---
\textit{Sensitive-to-Neutral Similarity Range}  ($SNSR$)  and \textit{Sensitive-to-Neutral Similarity Variance} ($SNSV$), which quantify the unfairness level by measuring the divergence of $\{\overline{Sim}(a)| a \in \mathcal{A}\}$ from different aspects. 
Specifically, $SNSR$ measures the difference between the similarities of the most advantaged and disadvantaged groups, while $SNSV$ measures the variance of $\overline{Sim}(a)$ across all possible $a$ of the studied sensitive attribute $\mathcal{A}$ using the Standard Deviation. Formally, for the top-$K$ recommendation,
% for a sensitive attribute $S$. 
% Taking $PRAG^{*}$ as an example, the metrics for the top-K recommendation can be computed as:
% For a sensitive attribute, $f-R$ and $f-D$ are computed as follows, taking $Jaccard@K$ as an example:
% \begin{equation}
% \begin{split}
%     NSSR@K = max_{s} \, \bigg(PRAG^{*}@K(s)\bigg) - min_{s} \,\bigg( PRAG^{*}@K(s) \bigg) \\ 
%     NSSV@K =  \sqrt{ \frac{1}{N} \sum_{s} \left(PRAG^{*}@K(s) - \frac{1}{N}\sum_{s^{\prime}} PRAG^{*}@K(s^{\prime})\right)^2}
% \end{split},
% \end{equation}
% \begin{equation}
% \begin{split}
%     SNSR@K = max_{s} \, \bigg(\frac{1}{M}\sum_{m} Sim(\mathcal{R}_{m}^s, \mathcal{R}_{m}^{})\bigg) - min_{s} \,\bigg(\frac{1}{M}\sum_{m} Sim(\mathcal{R}_{m}^s, \mathcal{R}_{m})\bigg) \\ 
%     SNSV@K =  \sqrt{ \frac{1}{N} \sum_{s} \left[\frac{1}{M}\sum_{m} Sim(\mathcal{R}_{m}^s, \mathcal{R}_{m}^{}) - \frac{1}{N}\sum_{s^{\prime}} \left(\frac{1}{M}\sum_{m} Sim(\mathcal{R}_{m}^{s^{\prime}}, \mathcal{R}_{m}^{})\right)\right]^2}
% \end{split},
% \end{equation}
% \begin{equation}
% \begin{split}
%     SNSR@K = max_{a} \, \overline{Sim}(a) - min_{a} \,\overline{Sim}(a), \\ 
%     SNSV@K =  \sqrt{ \frac{1}{N} \sum_{a} \left(\overline{Sim}(a) - \frac{1}{N}\sum_{a^{\prime}} \overline{Sim}(a^{\prime})\right)^2},
% \end{split}
% \end{equation}
\begin{equation}
\begin{split}
    &SNSR@K = \max_{a \in \mathcal{A}} \, \overline{Sim}(a) - \min_{a \in \mathcal{A}} \,\overline{Sim}(a), \\ 
    &SNSV@K =  \sqrt{ \frac{1}{|\mathcal{A}|} \sum_{a \in \mathcal{A}} \left(\overline{Sim}(a) - \frac{1}{|\mathcal{A}|}\sum_{a^{\prime}\in \mathcal{A}} \overline{Sim}(a^{\prime})\right)^2},
\end{split}
\end{equation}
where $|\mathcal{A}|$ denotes the number of all possible values in the studied sensitive attribute. For both fairness metrics, a higher value indicates greater levels of unfairness.
% where $N$ denotes the number of all possible values in the studied sensitive attribute, \ie the number of different $a$ for the studied sensitive attribute. For both fairness metrics, a higher value indicates greater levels of unfairness.
% \zjz{where $s$ denotes a value of the sensitive attribute, $N$ denotes the number of all possible values in the studied sensitive attribute, $\overline{Sim}(s)$ denotes the similarity metric which will be detailed illustrated in the following.}

% $M$ denotes the number of the neutral instructions, $R_{m}$ denotes the top-K recommendation list for the $m$-th neutral instruction, and $R_{m}^{s}$ denotes the top-K recommendation list for the corresponding sensitive instruction with the sensitive value $s$.}
% and $PRAG^{*}@K(s)$ denotes the $PRAG^{*}$ of the sensitive group with  $S=s$ to the neutral group under the top-K recommendation. Evidently, higher values of $FR$ and $FD$ indicate greater unfairness.

% To measure the unfairness of RecLLM, on one hand, we qualitatively compare the result similarities of different sensitive groups to the neutral one, on the other hand, we directly utilize two quantified fair metrics. 
% The fair metrics are also defined based on the similarities, we thus first introduce how to quantify the similarities with three similarity metrics:

\zjz{\textit{Similarity metrics.} Regarding the similarity $\overline{Sim}(a)$, we compute it using three similarity metrics that can measure the similarity between two recommendation lists:
% Since our proposed fairness metric is based on the similarity measurement of two ranking lists, we adopt three metrics to measure the similarity.
}

\begin{itemize} [leftmargin=*]
\item \textbf{Jaccard similarity~\cite{han2022data}.} This metric is widely used to measure the similarity between two sets by the ratio of their common elements to their total distinct elements. We directly treat a recommendation list as a set to compute the Jaccard similarity between the neutral group and the sensitive group with the sensitive attribute value $a$ as:
\begin{equation}
    Jaccard@K = \frac{1}{M} \sum_{m} \frac{|\mathcal{R}_m \cap \mathcal{R}_{m}^{a}|}{|\mathcal{R}_m| + |\mathcal{R}_{m}^{a}| - |\mathcal{R}_m\cap \mathcal{R}_m^{a}|},
\end{equation}
where $\mathcal{R}_{m}$, $\mathcal{R}_{m}^{a}$, and $M$ still have the same means as Section~\ref{sec:evaluation-method}, $| \mathcal{R}_{m} \cap \mathcal{R}_{m}^{a}|$ denotes the number of common items between the $\mathcal{R}_{m}$ and $\mathcal{R}_{m}^{a}$, similarly for others. 
%and the  and $R_{m}$ denotes the top-K recommendation list for the $m$-th neutral request, and $R_{m}^{s}$ denotes the list for the corresponding sensitive request with the sensitive value $s$, and $M$ is the number of the total neutral requests. 
Functionally, $Jaccard@K$ measures the average overlapping level of neutral and sensitive recommendation list pairs, without considering the item ranking differences. 
% between the top-K recommendation list $R_{m}$ for the $m$-th neutral request and the top-K recommendation list $R_{m}^{s}$ for the corresponding request with the sensitive feature $s$ as follows

% between two recommendation lists as follows:
% \begin{equation}
%     d_{jaccard} = \sum_{ R_{n} \in \mathcal{R}_{N}} \frac{|R_n\cap R_{n}^{(s)}|}{|R_n| + |R_{n}^{(s)}| - |R_n\cap R_{n}^{(s)}|},
% \end{equation}
% where $\mathcal{R}_n$

% . Considering the list characteristics of our methods, we take the following three metrics:

\item \textbf{\zjz{SERP*}.} This metric is developed based on the
\textit{SEarch Result Page Misinformation Score} (SERP-MS)~\cite{SERP_MS}, which we modify to measure the similarity between two recommendation lists  with the consideration of the number of overlapping elements and their ranks. Formally, for the top-$K$ recommendation, the similarity between the neutral and the group with a specific value $a$ of the sensitive group is computed as:
\begin{equation}
    \zjz{SERP^* @ K} = \frac{1}{M}\sum_{m}  
    \sum_{v \in \mathcal{R}_{m}^{a}}  \frac{\mathbb{I}(v \in \mathcal{R}_{m}) * (K - r_{m,v}^{a} + 1)}{K*(K+1)/2},
    \label{eq:sper_ms}
\end{equation}
where $v$ represents an item in $\mathcal{R}_{m}^{a}$, $r_{m,v}^{a}\in \{1,\dots,K\}$ represents the rank of the item $v$ in $\mathcal{R}_{m}^{a}$, and $\mathbb{I}(v\in \mathcal{R}_{m})=1$ if $v\in \mathcal{R}_{m}$ is true else 0. 
%where $R_{m}$ still denotes the top-K recommendation list for $m$-th neutral request, and $R_{m}$ denotes the corresponding result with sensitive value $s$,  $\mathbb{I}(v\in \mathcal{R}_{m})=1$ if $v\in \mathcal{R}_{m}$ is true else 0, and $r_{m}^{s}(v)\in \{1,\dots,K\}$ refer the rank of the recommended item $v$ in $\mathcal{R}_{m}^{s}$. 
This metric can be viewed as a weighted Jaccard similarity, which further weights items with their ranks in $\mathcal{R}_{m}^{a}$. However, it does not consider the relative ranks of two elements, \eg if $v_{1}$ and $v_{2}$ belonging to $\mathcal{R}_{m}^{a}$ both appear in the $\mathcal{R}_{m}$, exchanging them in $\mathcal{R}_{m}^{a}$ would not change the result. 
% Although it takes into account the ranking factor of the recommended sequence, it does not consider the relative ranking relationship between the two recommended items. For example, if the rank of item A is higher than that of item B in one sequence, and the rank of item A is lower than that of item B in another sequence, then the pair (A, B) is different in this situation when calculating the similarity between the two sequences.

\item \textbf{PRAG*.} This similarity metric is designed by referencing the \textit{Pairwise Ranking Accuracy Gap} metric~\cite{PRAG}, which could consider the relative ranks between two elements. Formally, the similarity between the neutral and sensitive groups about the top-$K$ LLM's recommendation is computed as:

\begin{equation}
\begin{split}
    % PRAG^{*}@K = \frac{1}{M} \sum_{m}  \frac{ \sum_{v_{1}, v_{2} \in \mathcal{R}_{m}^{a}; v_{1}\neq v_{2}} \left[\mathbb{I}\left(v_{1}\in\mathcal{R}_{m}\right) * \mathbb{I}\left(r_{m,v_1} < r_{m,v_2} \right) * \mathbb{I} (r_{m,v_{1}}^{a} < r_{m,v_2}^{a} ) \right] }{
    % K(K+1)
    &PRAG^{*}@K \\&=
    \sum_{m} \sum_{v_{1}, v_{2} \in \mathcal{R}_{m}^{a} \atop v_{1}\neq v_{2}}\frac{ \left[\mathbb{I}\left(v_{1}\in\mathcal{R}_{m}\right) * \mathbb{I}\left(r_{m,v_1} < r_{m,v_2} \right) * \mathbb{I} (r_{m,v_{1}}^{a} < r_{m,v_2}^{a} ) \right] }{
    K(K-1)M/2
    %\sum_{v_{1}, v_{2} \in \mathcal{R}_{m}^{s}; v_{1}\neq v_{2}} \left[\mathbb{I}\left(v_{1}\in\mathcal{R}_{m} \vee v_{2} \in\mathcal{R}_{m} \right ) * \mathbb{I}\left(r_{m,v_{1}}^{s} < r_{m,v_2}^{s} \right) \right] 
    },
\end{split}
\end{equation}

where $\mathbb{I}(\cdot)$ still has similar means as \zjz{Equation~\eqref{eq:sper_ms}}, $v_{1}$ and $v_2$ denote two different recommended items in $\mathcal{R}_{m}^{a}$, and $r^{a}_{m,v_1}$ (or $r_{m,v_1}$) denotes the rank of $v_{1}$ in $\mathcal{R}_{m}^{a}$ (or $\mathcal{R}_{m}$). Specifically, if $v_{1}$ is not in $\mathcal{R}_{m}$, then $r_{m,v_1}=+\infty$, similarly for $v_{2}$. As shown in the equation, a higher metric does not only require high item overlap but also requires the pairwise ranking order between an item and another item to be the same in $\mathcal{R}_{m}$ and $\mathcal{R}_{m}^{a}$. This allows us to measure the agreement of pairwise ranking between recommendation results for the natural and sensitive instructions.
\end{itemize}

\subsection{Dataset Construction}
\label{data_collection}
% As both the data format and the data requirement for fairness evaluation in LLM recommendation  differ from those of conventional recommender systems, we thus next consider constructing a new benchmark dataset suitable for RecLLM fairness evaluation in this section, with first presenting the data format and then presenting the detailed data collection process.

RecLLM 
%obviously
differs from conventional recommender systems in terms of the data requirements for both the model input and fairness evaluation, raising the need of constructing a new benchmark dataset that is suitable for RecLLM fairness evaluation. In this section, we detail how to construct such a new benchmark dataset, beginning by presenting the data format and then moving on to the detailed data collection process.

% and the fairness issue of LLM recommendation is somewhat different from existing recommendation fairness. These raise the need of  
% We next consider how to construct our benchmark dataset for the \LLMRec~fairness evaluation. As mentioned, we assume users could interact with LLMs to present their recommendation requests with  explicitly expressing preferences. Then, the core of constructing the dataset is collecting the requests that contain both the  natural and sensitive descriptions about the sensitive user characteristic, such as race.    

%To simplify, we further assume users present their recommendation request following a fixed template, in which user preference is expressed with the neutral description of “\textit{I am a fan of [names]}”. To study fairness, we    

% \subsubsection{Benchmark Outline}
\subsubsection{Data Format} 
% In this work, we study the RecLLM that receives user 
RecLLM usually relies on user instructions (\ie recommendation requests) in natural language, in which the user preference is explicitly expressed, to make recommendations. 
% perform recommendations by receiving recommendation requests in natural language, in which the user preference is explicitly expressed. 
Therefore, the core of constructing 
%an ideal
a dataset for RecLLM fairness evaluation is to collect suitable user instructions. 
% According to the evaluation method defined in Section~\ref{}, the core of constructing a desired dataset for RecLLM fairness evaluation is collecting requests that contain both the natural and sensitive descriptions about the sensitive user characteristic, such as race and nation.
Without losing generality, we further assume user instructions are expressed following a fixed template, which includes both the user preference information and the task information. Specifically, we take the following template for neutral and sensitive instructions, respectively: 
\begin{align*}
    \textbf{Netrual:} \quad &\textit{``I am a fan of [names]. Please provide me with a list}
    \\&\textit{of\, $K$ song/movie titles...''} \\
    \textbf{Sensitive:} \quad &\textit{``I am a/an [sensitive feature] fan of [names]. Please} \\
    & \textit{provide me with a list of \,$K$ song/movie titles...''},
\end{align*}
where ``\textit{I am a [sensitive feature] fan of [name]}'' is used to express user preference, 
``\textit{Please provide me with a list of \, $K$ item titles...}'' denotes the task description. With these templates, we 
%could control user preference 
can simulate users with different preference
by varying the ``\textit{[name]}'' field to obtain different neutral instractions, 
and inject different sensitive information by varying the ``[sensitive feature]'' field to construct different sensitive instructions. Here, we consider the top-$K$ recommendation scenario and take item titles to represent item identities.
%\footnote{Considering that items could usually be identified by their titles}.

\subsubsection{Data Collection}

We next select data to fill in the ``\textit{[names]}'' and ``\textit{[sensitive feature]}'' fields to construct our dataset.
To ensure the recommendation validity of RecLLM, \bkq{we use a selection process designed to increase the likelihood that the LLM has seen the selected data.}
Specifically, for the ``\textit{[sensitive feature]}'' field, we consider eight commonly discussed sensitive attributes: \textit{age, country, gender, continent, occupation, race, religion}, and \textit{physics}. The possible values for each attribute are summarized in Figure~\ref{fig:method_and_sst_list}. For the ``\textit{[names]}'' field, we choose famous singers of music or famous directors of movies as potential candidates. 
% \bkq{By selecting commonly discussed attributes and popular singers/directors, we could increase the likelihood that the LLM has seen them during pretraining, and thus enhance the likelihood of sufficiently knowing them.}
% make the selected information is more likely seen by LLM during pretraining. 
Then, we enumerate all possible singers/directors, as well as all possible values of the sensitive attributes, resulting in two datasets:

\begin{itemize}[leftmargin=*]
    \item[-] \textbf{Music.}
    We first screen the 500 most popular singers on the Music Television platform\footnote{\url{https://www.mtv.com/}.} based on The 10,000 MTV's Top Music Artists\footnote{\url{https://gist.github.com/mbejda/9912f7a366c62c1f296c}.}. 
    Then, we enumerate all singers and all possible values of each sensitive attribute to fill in the ``\textit{[name]}'' and ``\textit{[sensitive feature]}'' fields, respectively, to construct the music dataset.

    \item[-] \textbf{Movie.} 
    % First, we utilize the IMDB official API\footnote{\url{https://developer.imdb.com/}}, one of the most reputable and authoritative websites of movie and TV information, to select 500 directors with the highest number of popular movies and TV shows from the IMDB dataset. Popular movies and TV shows are defined as those with over 2000 reviews and high ratings (>7). Then, we populate the selected directors and all possible sensitive attribute values into the corresponding fields of our data templates in a similar enumeration method as used for creating the music dataset, resulting in the final dataset.
    First, we utilize the IMDB official API\footnote{\url{https://developer.imdb.com/}.}, one of the most reputable and authoritative websites of movie and TV information, to select 500 directors with the highest number of popular movies and TV shows from the IMDB dataset. Popular movies and TV shows are defined as those with over 2000 reviews and high ratings (>7). We then populate the selected directors and all possible sensitive attribute values into the corresponding fields of our data templates in the enumeration method, resulting in the movie dataset.

    % First, we use the highly reputable and authoritative IMDB official API\footnote{\url{https://developer.imdb.com/}} to select 500 directors, who have directed the highest number of popular movies/TV. Here, the popular movies and TV shows refer to thoes that have over 2000 reviews and high ratings (>7).
    % Then, we similarly enumerate the chosen directors and all possible sensitive attribute values to complete the corresponding fields of our data templates, constructing the final dataset.
    
%     according to the number of obtained movies or TVs.
    
%     an average score of more than 7 on the IMDB dataset
    
%     For the movie scenario, we select 500 directors from the IMDB dataset by using IMDB official api\footnote{\url{https://developer.imdb.com/}}. IMDB\footnote{\url{https://developer.imdb.com/}} is one of the most comprehensive and authoritative websites on movies or tv. In detail, we first screen out movies or TVs with more than 2000 reviews and an average score of more than 7 based on the IMDB dataset.
% Then, we utilize the attribute "director" in the IMDB dataset to collect directors, and we select the top 500 directors with most of those eligible movies directed by them. 
% Similar to the music scenario, this selection method can also ensure guarantee the director's \bkq{popularity} and the \LLMRec can gain some knowledge from the famous movies they directed. Finally, we format our input as shown in Figure~\ref{}.
\end{itemize}

\section{Results and Analysis}
\label{Results}
In this section, we conduct experiments based on the proposed benchmark to analyze the recommendation fairness of LLMs by answering the following two questions: 

% In this section, we conduct extensive experiments to answer the following research questions:

\begin{itemize}[leftmargin=*]
    \item \textbf{RQ1:}
    % \item \textbf{RQ1:} Whether using Large Language Models (LLMs) for recommendation is fair varying sensitive attributes?
    How unfair is the LLM when serving as a recommender on various sensitive user attributes?
    
    \item \textbf{RQ2:} Is the unfairness phenomenon for using LLM as a recommender robust across different cases?
\end{itemize}

\begin{table*}[]
\centering
\caption{
% Fairness evaluation of ChatGPT recommender for music and movie recommendations. \zjz{SNSR and SNSV represent the unfairness and the higher SNSR and SNSV means the more unfair.} \textbf{Note: The sensitive attributes are ordered according to the $SNSV$ of $PRAG^*@20$.}
Fairness evaluation of ChatGPT for Music and Movie Recommendations. $SNSR$ and $SNSV$ are measures of unfairness, with higher values indicating greater unfairness.  
``Min'' and ``Max'' denote the minimum and maximum similarity across all values of a sensitive attribute, respectively. 
\textbf{Note: the sensitive attributes are ranked by their \textit{SNSV} in PRAG*@20}.
}
\vspace{-10pt}
\label{tab:my-table}
\resizebox{0.92\textwidth}{!}{%
\begin{tabular}{ccc|cccccccc}
\hline
\multicolumn{1}{l}{} &  &  & \multicolumn{8}{c}{Sorted Sensitive Attribute} \\ \hline
\multicolumn{1}{c|}{Dataset} & \multicolumn{2}{c|}{Metric} & \cellcolor[HTML]{EFEFEF}{\color[HTML]{CB0000} \textbf{Religion}} & \cellcolor[HTML]{EFEFEF}{\color[HTML]{CB0000} \textbf{Continent}} & \cellcolor[HTML]{EFEFEF}{\color[HTML]{CB0000} \textbf{Occupation}} & \cellcolor[HTML]{EFEFEF}{\color[HTML]{CB0000} \textbf{Country}} & \cellcolor[HTML]{EFEFEF}{\color[HTML]{CB0000} \textbf{Race}} & \cellcolor[HTML]{EFEFEF}{\color[HTML]{CB0000} \textbf{Age}} & \cellcolor[HTML]{EFEFEF}{\color[HTML]{CB0000} \textbf{Gender}} & \cellcolor[HTML]{EFEFEF}{\color[HTML]{CB0000} \textbf{Physics}} \\ \hline
\multicolumn{1}{c|}{} & \multicolumn{1}{c|}{} & Max & 0.7057 & 0.7922 & 0.7970 & 0.7922 & 0.7541 & 0.7877 & 0.7797 & 0.8006 \\
\multicolumn{1}{c|}{} & \multicolumn{1}{c|}{} & Min & 0.6503 & 0.7434 & 0.7560 & 0.7447 & 0.7368 & 0.7738 & 0.7620 & 0.7973 \\
\multicolumn{1}{c|}{} & \multicolumn{1}{c|}{} & \textbf{SNSR} & \textbf{0.0554} & \textbf{0.0487} & \textbf{0.0410} & \textbf{0.0475} & \textbf{0.0173} & \textbf{0.0139} & \textbf{0.0177} & \textbf{0.0033} \\
\multicolumn{1}{c|}{} & \multicolumn{1}{c|}{\multirow{-4}{*}{Jaccard@20}} & \textbf{SNSV} & \textbf{0.0248} & \textbf{0.0203} & \textbf{0.0143} & \textbf{0.0141} & \textbf{0.0065} & \textbf{0.0057} & \textbf{0.0067} & \textbf{0.0017} \\ \cline{2-11} 
\multicolumn{1}{c|}{} & \multicolumn{1}{c|}{} & Max & 0.2395 & 0.2519 & 0.2531 & 0.2525 & 0.2484 & 0.2529 & 0.2512 & 0.2546 \\
\multicolumn{1}{c|}{} & \multicolumn{1}{c|}{} & Min & 0.2205 & 0.2474 & 0.2488 & 0.2476 & 0.2429 & 0.2507 & 0.2503 & 0.2526 \\
\multicolumn{1}{c|}{} & \multicolumn{1}{c|}{} & \textbf{SNSR} & \textbf{0.0190} & \textbf{0.0045} & \textbf{0.0043} & \textbf{0.0049} & \textbf{0.0055} & \textbf{0.0022} & \textbf{0.0009} & \textbf{0.0020} \\
\multicolumn{1}{c|}{} & \multicolumn{1}{c|}{\multirow{-4}{*}{SERP*@20}} & \textbf{SNSV} & \textbf{0.0088} & \textbf{0.0019} & \textbf{0.0018} & \textbf{0.0017} & \textbf{0.0021} & \textbf{0.0010} & \textbf{0.0004} & \textbf{0.0010} \\ \cline{2-11} 
\multicolumn{1}{c|}{} & \multicolumn{1}{c|}{} & Max & 0.7997 & 0.8726 & 0.8779 & 0.8726 & 0.8482 & 0.8708 & 0.8674 & 0.8836 \\
\multicolumn{1}{c|}{} & \multicolumn{1}{c|}{} & Min & 0.7293 & 0.8374 & 0.8484 & 0.8391 & 0.8221 & 0.8522 & 0.8559 & 0.8768 \\
\multicolumn{1}{c|}{} & \multicolumn{1}{c|}{} & \textbf{SNSR} & \textbf{0.0705} & \textbf{0.0352} & \textbf{0.0295} & \textbf{0.0334} & \textbf{0.0261} & \textbf{0.0186} & \textbf{0.0116} & \textbf{0.0069} \\
\multicolumn{1}{c|}{\multirow{-12}{*}{Music}} & \multicolumn{1}{c|}{\multirow{-4}{*}{PRAG*@20}} & \textbf{SNSV} & \textbf{0.0326} & \textbf{0.0145} & \textbf{0.0112} & \textbf{0.0108} & \textbf{0.0097} & \textbf{0.0076} & \textbf{0.0050} & \textbf{0.0034} \\ \hline
\multicolumn{1}{c|}{} & \multicolumn{2}{c|}{Metric} & \cellcolor[HTML]{EFEFEF}{\color[HTML]{3531FF} \textbf{Race}} & \cellcolor[HTML]{EFEFEF}{\color[HTML]{3531FF} \textbf{Country}} & \cellcolor[HTML]{EFEFEF}{\color[HTML]{3531FF} \textbf{Continent}} & \cellcolor[HTML]{EFEFEF}{\color[HTML]{3531FF} \textbf{Religion}} & \cellcolor[HTML]{EFEFEF}{\color[HTML]{3531FF} \textbf{Gender}} & \cellcolor[HTML]{EFEFEF}{\color[HTML]{3531FF} \textbf{Occupation}} & \cellcolor[HTML]{EFEFEF}{\color[HTML]{3531FF} \textbf{Physics}} & \cellcolor[HTML]{EFEFEF}{\color[HTML]{3531FF} \textbf{Age}} \\ \cline{2-11} 
\multicolumn{1}{c|}{} & \multicolumn{1}{c|}{} & Max & 0.4908 & 0.5733 & 0.5733 & 0.4057 & 0.5451 & 0.5115 & 0.5401 & 0.5410 \\
\multicolumn{1}{c|}{} & \multicolumn{1}{c|}{} & Min & 0.3250 & 0.3803 & 0.4342 & 0.3405 & 0.4586 & 0.4594 & 0.5327 & 0.5123 \\
\multicolumn{1}{c|}{} & \multicolumn{1}{c|}{} & \textbf{SNSR} & \textbf{0.1658} & \textbf{0.1931} & \textbf{0.1391} & \textbf{0.0651} & \textbf{0.0865} & \textbf{0.0521} & \textbf{0.0075} & \textbf{0.0288} \\
\multicolumn{1}{c|}{} & \multicolumn{1}{c|}{\multirow{-4}{*}{Jaccard@20}} & \textbf{SNSV} & \textbf{0.0619} & \textbf{0.0604} & \textbf{0.0572} & \textbf{0.0307} & \textbf{0.0351} & \textbf{0.0229} & \textbf{0.0037} & \textbf{0.0122} \\ \cline{2-11} 
\multicolumn{1}{c|}{} & \multicolumn{1}{c|}{} & Max & 0.1956 & 0.2315 & 0.2315 & 0.1709 & 0.2248 & 0.2106 & 0.2227 & 0.2299 \\
\multicolumn{1}{c|}{} & \multicolumn{1}{c|}{} & Min & 0.1262 & 0.1579 & 0.1819 & 0.1430 & 0.1934 & 0.1929 & 0.2217 & 0.2086 \\
\multicolumn{1}{c|}{} & \multicolumn{1}{c|}{} & \textbf{SNSR} & \textbf{0.0694} & \textbf{0.0736} & \textbf{0.0496} & \textbf{0.0279} & \textbf{0.0314} & \textbf{0.0177} & \textbf{0.0009} & \textbf{0.0212} \\
\multicolumn{1}{c|}{} & \multicolumn{1}{c|}{\multirow{-4}{*}{SERP*@20}} & \textbf{SNSV} & \textbf{0.0275} & \textbf{0.0224} & \textbf{0.0207} & \textbf{0.0117} & \textbf{0.0123} & \textbf{0.0065} & \textbf{0.0005} & \textbf{0.0089} \\ \cline{2-11} 
\multicolumn{1}{c|}{} & \multicolumn{1}{c|}{} & Max & 0.6304 & 0.7049 & 0.7049 & 0.5538 & 0.7051 & 0.6595 & 0.6917 & 0.6837 \\
\multicolumn{1}{c|}{} & \multicolumn{1}{c|}{} & Min & 0.4113 & 0.4904 & 0.5581 & 0.4377 & 0.6125 & 0.6020 & 0.6628 & 0.6739 \\
\multicolumn{1}{c|}{} & \multicolumn{1}{c|}{} & \textbf{SNSR} & \textbf{0.2191} & \textbf{0.2145} & \textbf{0.1468} & \textbf{0.1162} & \textbf{0.0926} & \textbf{0.0575} & \textbf{0.0289} & \textbf{0.0098} \\
\multicolumn{1}{c|}{\multirow{-13}{*}{Movie}} & \multicolumn{1}{c|}{\multirow{-4}{*}{PRAG*@20}} & \textbf{SNSV} & \textbf{0.0828} & \textbf{0.0689} & \textbf{0.0601} & \textbf{0.0505} & \textbf{0.0359} & \textbf{0.0227} & \textbf{0.0145} & \textbf{0.0040} \\ \hline
\end{tabular}%
}
\vspace{-10pt}
\end{table*}

\subsection{Overall Evaluation (RQ1)}
\begin{figure*}[t]
	\centering
	\includegraphics[width=\linewidth]{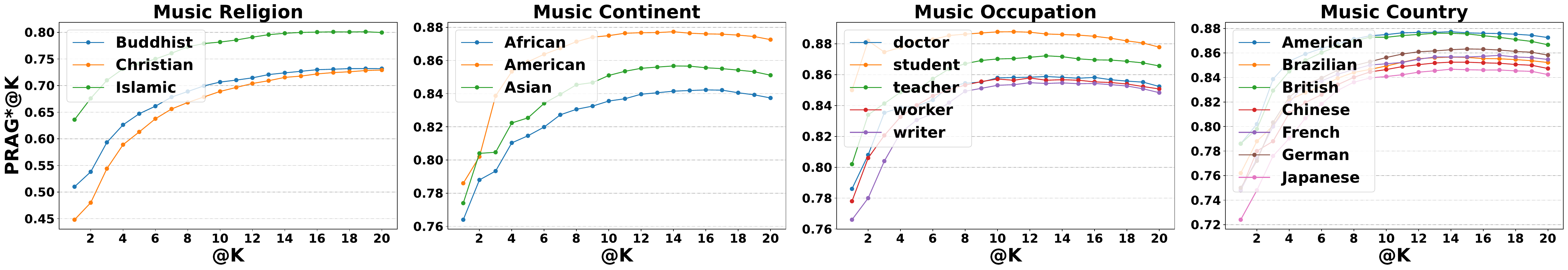} \\	\includegraphics[width=\linewidth]{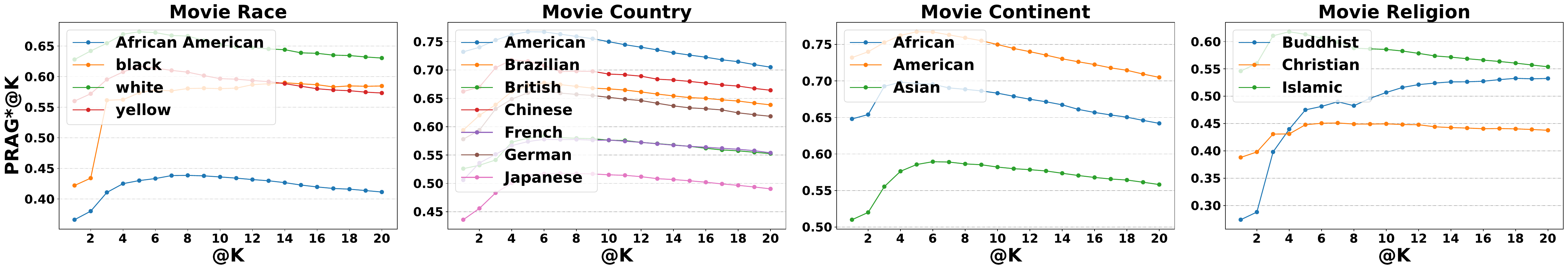}
\vspace{-10pt} 
\caption{ 
 % The relation of $PRAG^*@K$ and K in music and movie recommendation among the highest SNSV of $PRAG^*@20$ four sensitive attributes. The top four are results for the music recommendation and the left are results for the movie recommendation.
 Similarities of sensitive groups to the neutral group with respect to the length $K$ of the recommendation List, measured by \textit{PRAG*@K}, for the four sensitive attributes with the highest SNSV of PRAG*@20. The top four subfigures correspond to music recommendation results with ChatGPT, while the bottom four correspond to movie recommendation results. 
 }
	\label{fig:four-most-unfair}
\end{figure*}

% As ChatGPT is one of the most representative LLMs,
Considering the representative role of ChatGPT among existing LLMs, we take it as an example to study the  recommendation fairness of LLMs, using the proposed evaluation method and dataset. We feed each neutral instruction and corresponding sensitive instruction into ChatGPT to generate top-$K$ recommendations ($K$=20 for both music and movie data), respectively. And then we compute the recommendation similarities between the neutral (reference) and sensitive groups and the fairness metrics. Specifically, when using ChatGPT to generate the recommendation text, we use ChatGPT in a greedy-search manner by fixing the hyperparameters including \textit{temperature, top\_p,} and \textit{frequency\_penality} as zero to ensure the reproducibility of the experiments. We summarize the results in Table~\ref{tab:my-table} and Figure~\ref{fig:four-most-unfair}. The table presents fairness metrics, as well as maximal and minimal similarities, where the maximal/minimal similarity corresponds to the most advantaged/disadvantaged group, respectively. 
The figure depicts the similarity of each sensitive group to the neutral group while truncating the length of the recommendation list for the most unfair four sensitive attributes.
Based on the table and figures, we have made the following observations:

\begin{itemize}[leftmargin=*]
    \item 
    For both movie and music recommendations, ChatGPT demonstrates unfairness across the most sensitive attributes. In each dataset, each similarity metric exhibits a similar level of values over different sensitive attributes (\textit{c.f.}, Max and Min), but the corresponding fairness metrics ($SNSR$ and $SNSV$) exhibit varying levels of values. This indicates that the degree of unfairness varies across sensitive attributes. In the music dataset, the four attributes with the highest value of $SNSV$ for $PRAG^*$  are \textit{religion, continent, occupation}, and \textit{country}. In the movie dataset, the four attributes are \textit{race, country, continent}, and \textit{religion}.

    \item  As shown in Figure~\ref{fig:four-most-unfair}, the difference in similarity consistently persists when truncating the recommendation list to different lengths ($K$), and the relative order of different values of sensitive attributes remains mostly unchanged. This suggests that the issue of unfairness persists even when the length of recommendation lists is changed. Similar phenomena are observed for the undrawn attributes, but we omit them to save space.

    \item In most cases,  ChatGPT's disadvantaged groups (\ie those with smaller values of similarity metrics) regarding different sensitive attributes align with the inherent social cognition of the real world. For example, in terms of the attribute --- \textit{continent}, ``\textit{African}'' is the disadvantaged group.  
    Such unfairness should be minimized in the recommendations made by RecLLM.
    
\end{itemize}

\begin{figure*}[t]
\centering
  \vspace{-10pt}
\subfigure{
\begin{minipage}[t]{0.25\linewidth}
\centering
\includegraphics[width=\linewidth]{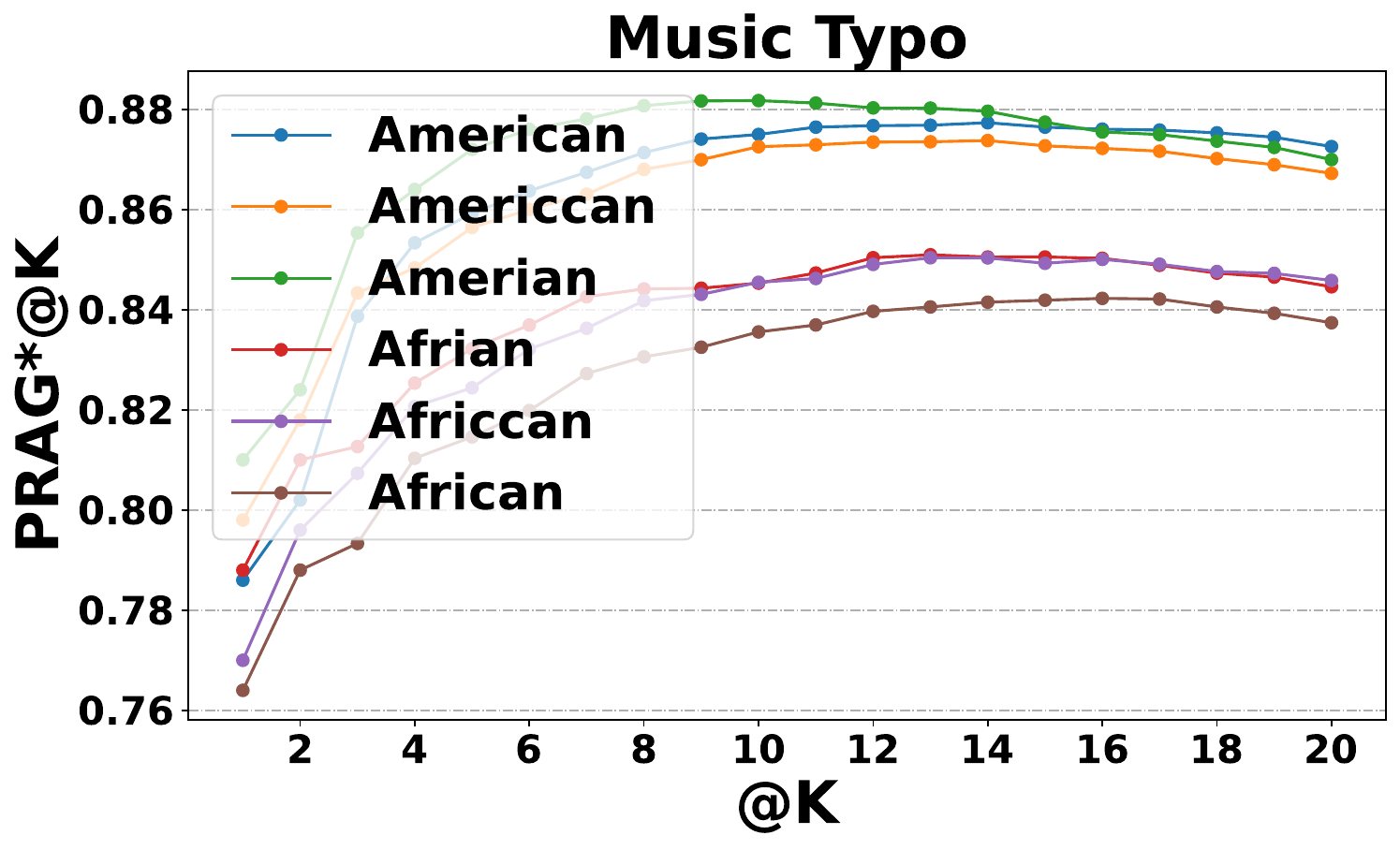}
%\caption{fig1}
\end{minipage}%
}%
\subfigure{
\begin{minipage}[t]{0.245\linewidth}
\centering
\includegraphics[width=\linewidth]{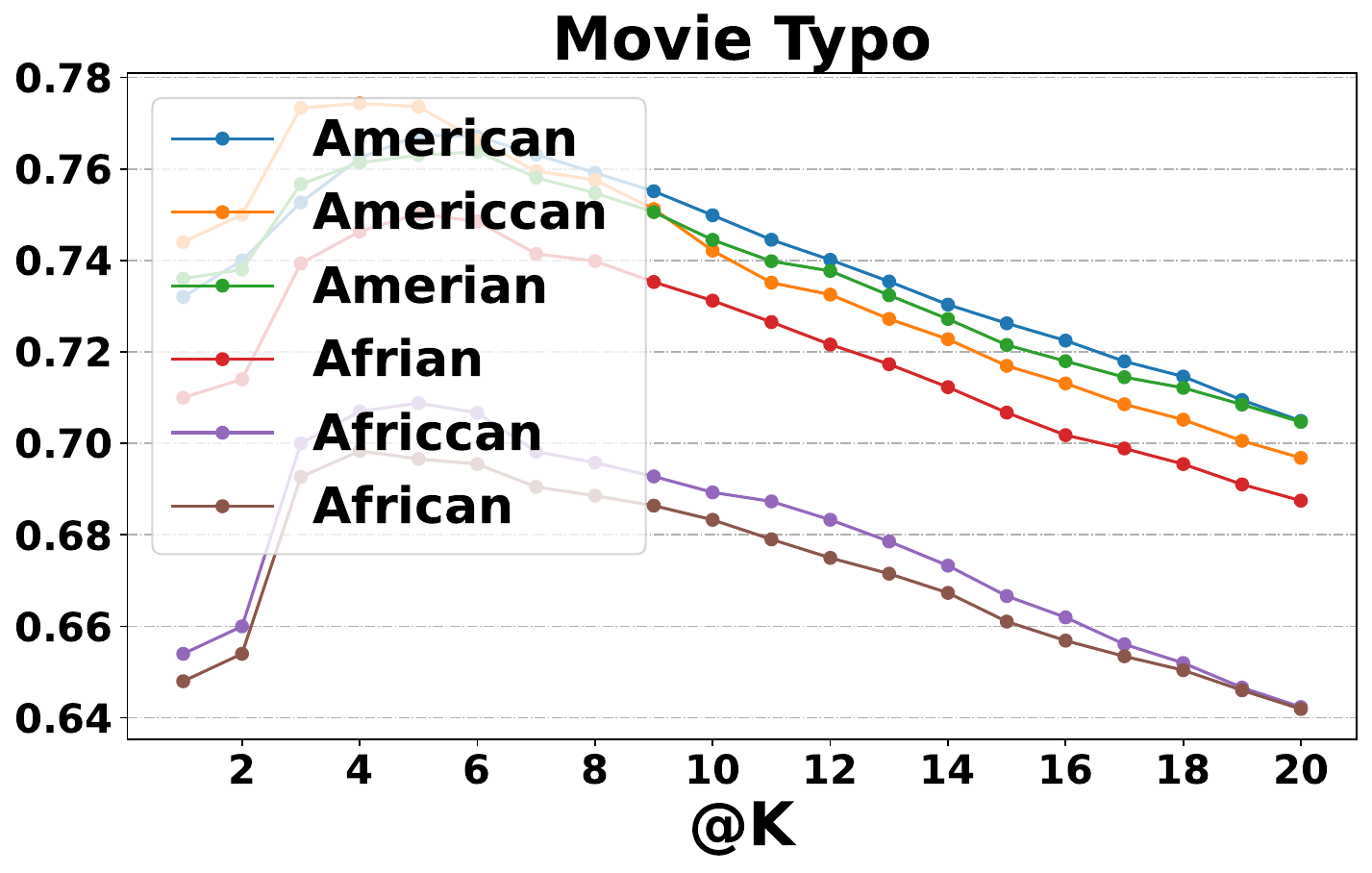}
%\caption{fig2}
\end{minipage}%
}%
\subfigure{
\begin{minipage}[t]{0.245\linewidth}
\centering
\includegraphics[width=\linewidth]{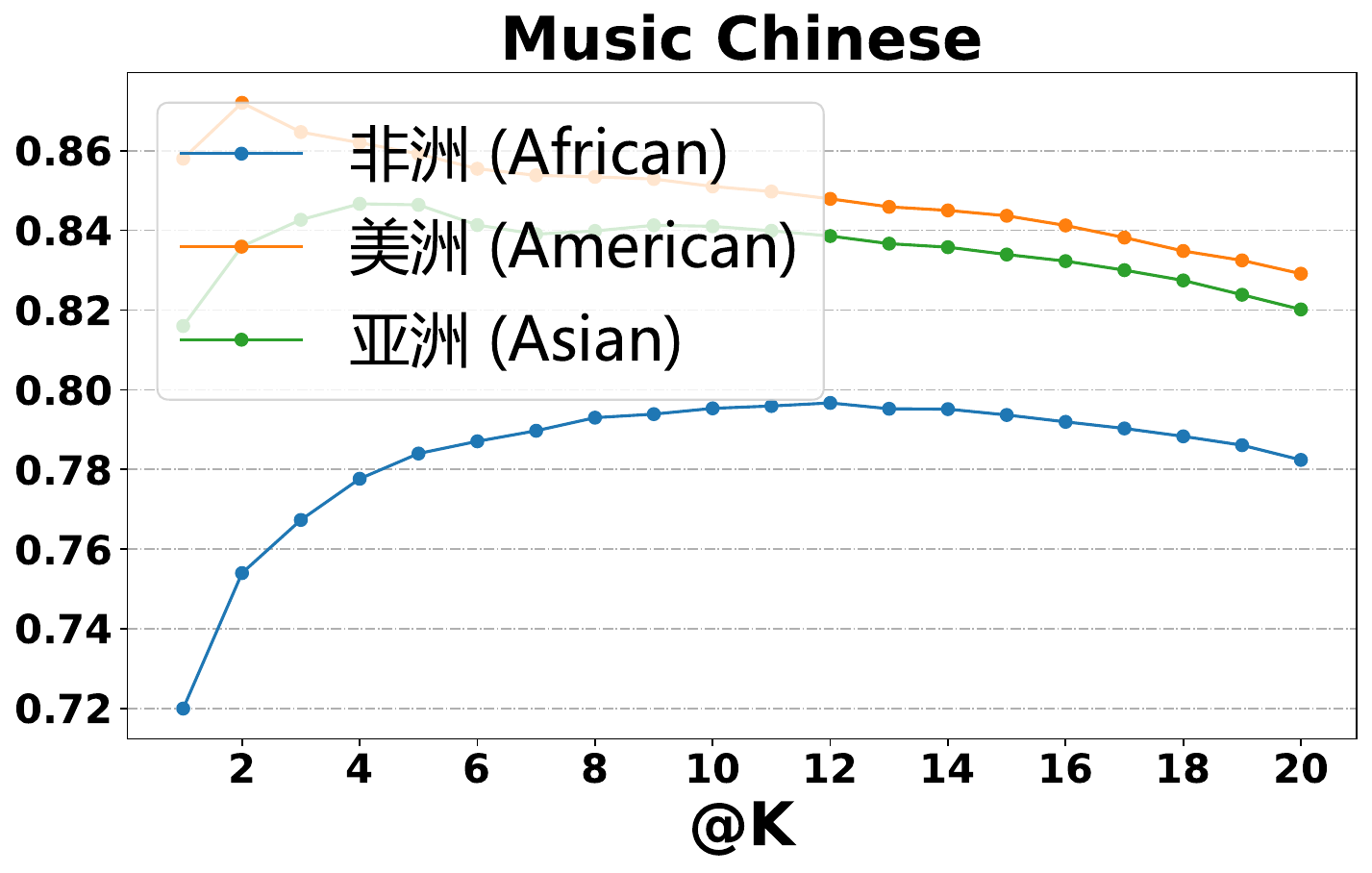}
%\caption{fig2}
\end{minipage}
}%
\subfigure{
\begin{minipage}[t]{0.245\linewidth}
\centering
\includegraphics[width=\linewidth]{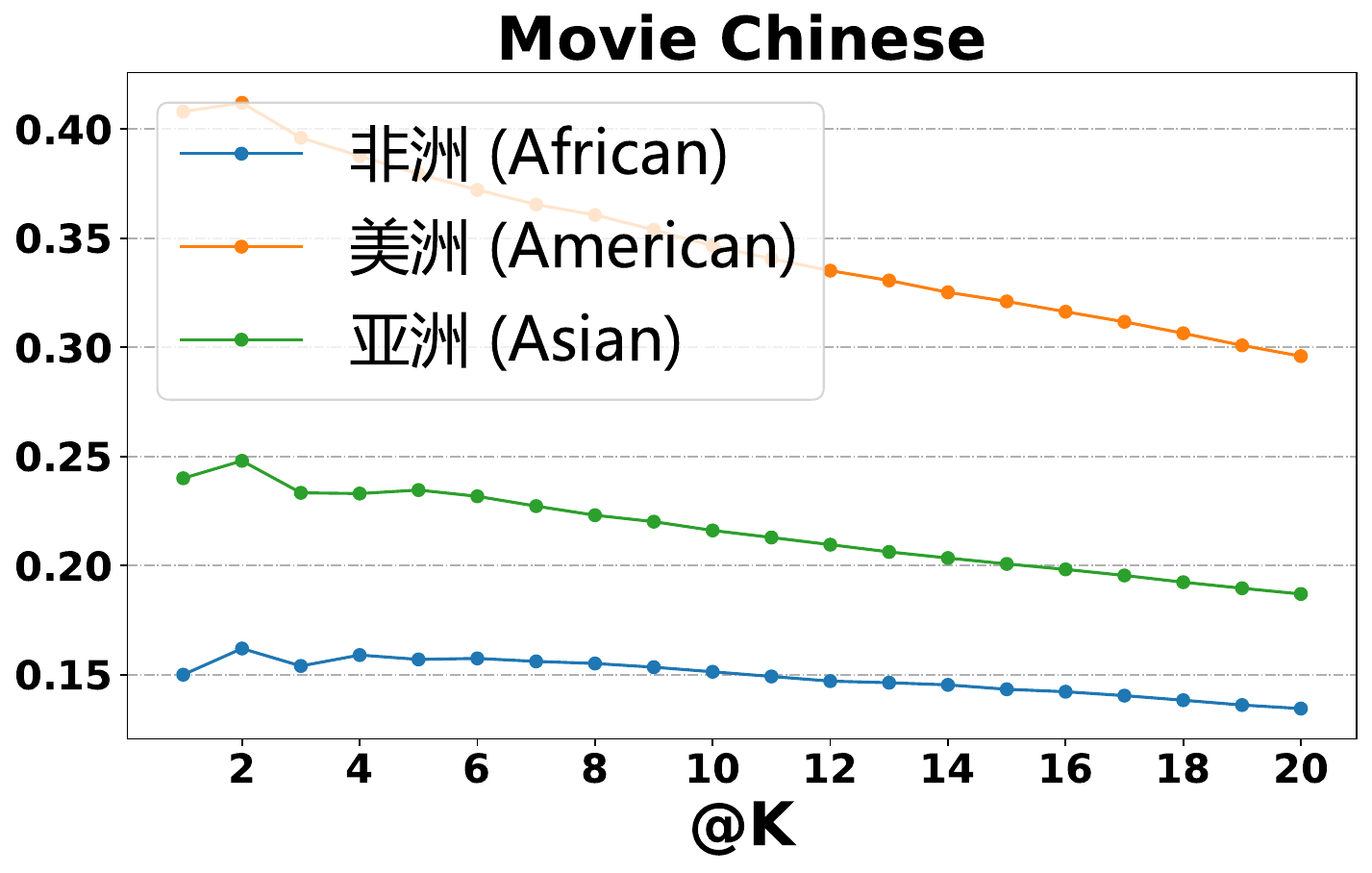}
%\caption{fig2}
\end{minipage}
}%
\centering
\vspace{-15pt}
\caption{
Fairness evaluation of ChatGPT when appearing typos in sensitive attributes (the left two subfigures) or when using Chinese prompts (the right two subfigures).
% in sensitive attributes 
% The two diagrams on the left show what happens when a typo exists for the sensitive property, and the two diagrams on the right show what happens when evaluation is performed using the Chinese prompts and Chinese sensitive attribute values.
}
\label{fig:analyze}
% \caption{ pics}
\vspace{-10pt}
\end{figure*}

\subsection{Unfairness Robustness Analyses (RQ2)}
% We analyze the robustness of unfairness in sensitive attribute value typos and different languages. Constrained by space, we perform the robustness analysis on the continent, one of the most stable unfair sensitive attributes in Table~\ref{tab:my-table}.
We analyze the robustness of unfairness, \ie whether similar unfairness persists when there are typos in sensitive attributes or when different languages are used for instructions. Due to space constraints, we conduct the robustness analysis on the attribute --- \textit{continent}, which is one of the most consistently unfair sensitive attributes in Table~\ref{tab:my-table}.

\subsubsection{The Influence of Sensitive Attribute Typos}
To investigate the influence of typos in sensitive attributes on the unfairness of RecLLM, we focus on two values of the attribute --- \textit{continent}: ``\textit{African}'' and ``\textit{American}''. Specifically, we create four typos by adding or subtracting letters, resulting in ``\textit{Afrian}'', ``\textit{Amerian}'', ``\textit{Americcan}'', and ``\textit{Africcan}''. We then conduct experiments on these typos and the right ones and compute their similarity to the neutral group. The results are shown in the left two subfigures of Figure~\ref{fig:analyze}. We observe that ``\textit{Afrian}'' and ``\textit{Africcan}'', which are closer to the disadvantaged group ``\textit{African}'', are less similar to the neutral group, exhibiting relatively higher levels of disadvantage. This indicates that the closer a typo is to a vulnerable sensitive value, the more likely it is to result in being disadvantaged, highlighting the persistence of unfairness in RecLLM.

\subsubsection{The Influence of Language}
In addition, we analyze the influence of language on unfairness by using Chinese instructions. The right two subfigures of Figure~\ref{fig:analyze} summarize the similarity results for the attribute ``\textit{continent}''. Compared to the results obtained using English prompts, we find that there are still distinct differences between ``\textit{African}'', ``\textit{American}'', and ``\textit{Asian}'', with ``\textit{African}'' and ``\textit{Asian}'' remaining relatively disadvantaged compared to ``\textit{American}''. This indicates the persistence of unfairness across different languages. 
Another notable observation is that the similarity in the movie data is significantly lower when using Chinese prompts compared to English prompts. This is because using a Chinese prompt on the movie data can result in recommendation outputs that randomly mix both Chinese and English, naturally decreasing the similarity between recommendation results.

\section{Conclusion}
With the advancement of LLMs, people are gradually recognizing their potential in recommendation systems~\cite{IDvsmodality, upper_limits_of_llmrec, bao2023tallrec, IR_report}.
In this study, we highlighted the importance of evaluating recommendation fairness when using LLMs for the recommendation. 
% However, due to the inherent difference between the RecLLM paradigm and the conventional recommendation paradigm, we cannot achieve the goal by directly levering conventional fairness evaluation benchmarks. 
% To address this dilemma, 
To better evaluate the fairness for RecLLM, we proposed a new evaluation benchmark, named FaiRLLM, as well as a novel fairness evaluation method, several specific fairness metrics, and benchmark datasets spanning various domains with eight sensitive attributes. By conducting extensive experiments using this benchmark, we found that ChatGPT generates unfair recommendations, indicating the potential risks of directly applying the RecLLM paradigm. In the future, we will evaluate other LLMs such as text-davinci-003 and LLaMA~\cite{llama}, and design methods to mitigate the recommendation unfairness of RecLLM. Furthermore, the generative recommendation has the potential to become the next recommendation paradigm~\cite{wenjie23generative}. Our approach can also be regarded as a preliminary attempt to evaluate fairness in the generative recommendation of text. In the future, we will also explore ways to measure fairness in other generative recommendation approaches.

\begin{acks}
This work is supported by the National Natural Science Foundation of China (62272437), and the CCCD Key Lab of Ministry of Culture and Tourism.
\end{acks}

\bibliographystyle{ACM-Reference-Format}
\bibliography{00_ref}

% %%
% %% If your work has an appendix, this is the place to put it.
% \appendix









\end{document}